\documentclass[twocolumn,numbering,showpacs]{revtex4-1}

\usepackage{graphicx,color}
\usepackage{latexsym}
\usepackage{amsmath,amssymb}        
\usepackage[draft=false]{hyperref}
\usepackage{mathrsfs}
\usepackage{comment}
\usepackage{soul}
\usepackage{mathrsfs}
\definecolor{purple}{rgb}{0.58,0.0,0.83}
\definecolor{orange}{rgb}{1,0.5,0}
\DeclareSymbolFontAlphabet{\mathrsfs}{rsfs}
\DeclareMathAlphabet{\mathcal}{OMS}{cmsy}{m}{n}

\begin{document}

% -----> TITLE 

\title{Head-on collision of multi-state ultralight BEC dark matter configurations}

% ----->   AUTHORS   <-----

\author{F. S. Guzm\'an${}^{1}$ and Ana A. Avilez${}^{2}$}
\affiliation{${}^{1}$ Laboratorio de Inteligencia Artificial y Superc\'omputo,
	      Instituto de F\'{\i}sica y Matem\'{a}ticas, Universidad
              Michoacana de San Nicol\'as de Hidalgo. Edificio C-3, Cd.
              Universitaria, 58040 Morelia, Michoac\'{a}n,
              M\'{e}xico.\\
              ${}^{2}$ Departamento de F\'{\i}sica,
  Centro de Investigaci\'on y de Estudios Avanzados del IPN,\\
  A. P. 14--740,  07000, Ciudad de M\'exico, M\'exico.}

% --->   DATE

\date{\today}

% -----> ABSTRACT

\begin{abstract}
Density profiles of ultralight Bose-Condensate dark matter inferred from numerical simulations of structure formation, ruled by the Gross-Pitaevskii-Poisson (GPP) system of equations, have a core-tail structure. Multi-state equilibrium configurations of the GPP system on the other hand, have a similar core-tail density profile. We now submit these multi-state configurations to highly dynamical scenarios and show their potential as providers of appropriate density profiles of structures. What we do is to present the 
simulation of head-on collisions between two equilibrium configurations of the GPP system of equations, including the collision of ground state with multi-state configurations. We study the regimes of solitonic and merger behavior, and show generic properties of the dynamics of the system, including the relaxation process and attractor density profiles. We show the merger of multi-state configurations have the potential to produce core-tail density profiles, with the core dominated by the ground state and a halo dominated by an additional states.
\end{abstract}

% ----->   PACS

\pacs{keywords: dark matter -- Bose condensates}
%07.05.Tp Computer modeling and simulation
%07.05.Mh Neural networks, fuzzy logic, artificial intelligence
%05.45.Tp Time series analysis
%04.30.-w Gravitational waves

% ----->   MAKETITLE   <-----

\maketitle

% --------------------------------------------
% ----->     INTRODUCTION     <-----
% --------------------------------------------
\section{Introduction}
\label{sec:introduction}

Scalar fields have been essential in modern physics, at fundamental level the Higgs scalar that is responsible for the mass of elementary particles according to the standard model of particles, is  the most remarkable example. However, scalar fields have also been quite popular in modeling new forms of matter and energy whose existence manifests at astronomical and cosmological scales. Particularly, they have been used as effective field theories of dark matter and dark energy. Concerning dark matter in particular, 
a common application has been the modeling of astrophysical sized objects as classical solutions of a canonical scalar field holding a wave-length comparable to stellar scales or larger \citep{Kaup:1968,Ruffini:1969,Ruffini:1983,Chavanis:2007}. 
In particular at galactic sizes with the aim of proposing a model of dark matter understood as condensates \citep{Sin:1994,Ji:1994} and studied as classical scalar fields as well, with an interpretation of an effective theory \cite{Matos-Guzman:1998}. From the cosmological scenario it was found that the boson mass had to be ultralight in order to reproduce the observed distribution of matter at very large scales\citep{Matos-Urena:2000,Sahni:1999}. Similar proposals relying on the previous notion of haloes as huge condensates or classical scalar field configurations was reconsidered several times and different models arose under different denominations \citep{Hu:2000,GuzmanUrena2006,Harko:2007}. In recent years, the support of these models from a cosmological point of view has been enhanced and the interpretation as a Bose Condensate has become more suitable \citep{Boehmer:2007um,Chavanis-Harko:2011,Marsh-Ferreira:2010,Ostriker:2016,Chavanis:2016ial}.

The dynamics of the system of equations suitable for scalar fields and Bose Condensates were first analyzed in the nineties, using the tools of numerical relativity applied to the solution of the evolution equations, both for complex scalar fields with stationary solutions and real scalar fields with non-stationary solutions \citep{Seidel-Suen:1990,Seidel-Suen:1991,Seidel-Suen:1994}. These pioneering works showed the stability properties and  shed light on the existence and stability conditions of so called solitonic solutions in full General Relativity and in the linearized weak-field regime, known as the Schroedinger-Poisson (SP) system. A  relaxation mechanism called gravitational cooling was proposed, that  later was applied to galactic sized structures in order to provide a structure formation scenario after the turn-around point \citep{GuzmanUrena2003,GuzmanUrena2006,Chavanis:2017loo}. Ever since, with the motivation of dark matter and a relaxation process, more modern approaches suggested that the equations of the SP system are associated to a macroscopic Bose gas making up a Bose-Einstein condensate (BEC), both fully dynamical \citep{GuzmanUrena2006} or in limits like the Thomas-Fermi limit \cite{Harko:2007}.

Under this approach, after condensation, most of the individual bosons lay in the ground state and the macroscopic wave function scales as the occupation number of the BEC. In the mean-field approximation, the macroscopic wave function of such a system is described by the non-linear Gross-Pitaevskii equation  
\citep{Gross:1963,Pitaevskii:1961, Pitaevskii:2003}. In this way, the interpretation of the formerly SP system, becomes the Gross-Pitaevskii equation  for a Bose Condensate with a  potential trap that happens to be a gravitational potential sourced by the gas density itself. An important feature of this approach to model dark matter, is that the boson requires as an ultralight mass of order $m\sim 10^{-22} eV$, which suffices to suppress the formation of small structures, implying that a cut-off in the halo mass function and the matter power spectrum. Thus in this type of models the missing satellites is not a problem  \cite{Matos-Guzman:1998,Hu:2000,Suarez-review:2014,Ostriker:2016,Marsh-Ferreira:2010,Hlozek:2014,Schive:2015}. In addition, simulations of mergers of pure substructures as BECS at galactic and cluster scales, show that the equilibrium density profiles of this sort of dark matter have a core-like shape which is highly convenient in order to naturally ease the core-cusp controversy \citep{Chen:2016,Schive:2015,Du:2016,Robles:2013,Bernal:2016}. The core is compared to the solitonic solution resulting from the stationary solution of the GPP system  \cite{GuzmanUrena2004,GuzmanUrena2006}, which has a convenient set of properties, chief among them that these are attractor solutions for rather arbitrary initial configurations \cite{GuzmanUrena2006,BernalGuzman2006a}.

%\item Specific state of art regarding to cosmology and structure formation. 
In a cosmological context, the linear cosmological perturbations theory along with the background cosmology arisen from this model --plus a cosmological constant-- reproduces the observations at very large scales such as the linear matter power spectrum of the distribution of matter in the Universe and the CMB among others \citep{RodriguezMontoya:2010,suarez:2011,magana:2012,Urena-Lopez:2015gur,Cedeno:2017}. 
Beyond the linear cosmological regime,  N-body simulations in three dimensions, show that predictions of the matter power spectrum from CDM and SFDM  at cluster and supercluster scales at the non-linear cosmological threshold are indistinguishable. However, at galactic and intergalactic scales the story changes, because the density profiles of dark matter in the two scenarios differs importantly \citep{Schive:2014,Schive:2014hza,Schive:2015,Nori:2018}. 
In this direction, recent simulations of mergers of solitonic solutions of the Schrodinger-Poisson system  without self-interactions, indicate the final density profile have a central solitonic core density profile surrounded by a Navarro-Frenk-White(NFW)-like density tail \citep{Schwabe:2016,Schive:2014hza}. Although these simulations have improved substantially our understanding of the structure formation in BEC dark matter, a long track needs to be walked in order to realistically simulate the real structure formation. So far, in all the mentioned works, it is implicitly assumed that the distribution of the BEC is described only by a single wave function and it has been shown thoroughly that configurations in this state, when sufficiently isolated, they should approach a ground state configuration as long as the only relaxation mechanism is the gravitational cooling \cite{GuzmanUrena2006}.

Exploring further out single wave function systems, it was found that unlike ground state configurations, multi-state configurations show the property of having a core and a tail naturally \cite{UrenaBernal:2010,Bernal:2016}. We now explore the dynamics of this type of configurations in non-stationary scenarios. What we do in this paper is to simulate the head-on collision of two multi-state configurations and explore first, how the states deform due to the interaction with another structure and second, the possibilities that a multi-state scenario has at explaining the profile of structures resulting from binary mergers. We explore for the first time the head-on encounter of configurations initially laying not only in ground state but in different combinations of multi-states. Specifically, we simulate two types of collisions: the solitonic regime (or high velocity case) and the merger regime (or low velocity case). In the first regime, the magnitude of the head-on momenta is large enough to compensate the gravitational binding energy.  On the contrary, systems within the merger regime are bounded. In the two regimes we track the evolution process and simulate various representative scenarios, and for each one we present particular information revealing the main features of the dynamics.

Finally, we study two representative merging cases, the collision of a ground state with another ground state  and the collision of a two-states configurations with another two-states configurations. We study the relaxation process and the density profile of the final configurations in order to sketch the potential properties in a more general scenario of structure formation in the sense that we fit the density profile of the final configuration.

The paper is organized as follows. In Section \ref{sec:methods} we describe the GPP system of equations and the methods used to solve them. In Section \ref{sec:tests} we present the tests for the code used and in Section \ref{sec:results} we present the results of our analysis. Finally in Section \ref{sec:final} we present some conclusions and describe the possibilities of the multi-state scenario in the frame of structure profiles.

% -------------------------------------------
% ----->     Section     <-----
% ----------------bla---------------------------
\section{Methods}
\label{sec:methods}

% -----------------    Subsection
\subsection{Solution of the GPP system}

The GPP system of equations represents the description of a Bose Condensate in the regime at which the GP approach holds. A particular sign of the system is that the condensate trap potential of the GP equation corresponds to a gravitational potential sourced by the density of the wave function of the condensate, interpreted as a matter density \footnote{ Mentioned in the introduction, the classical solution of SP or the wave function arisen from GP scales as the occupation number of the condensate (or the groud state) since the wave-functions of individual particles are identical and therefore additive.}. The regular GP equation includes the self-interaction among bosons that has effects on the distribution of matter \cite{GuzmanUrena2006}, nevertheless in most of current applications to the ultralight dark matter model, the condensates are  described by the GPP in the free field case with zero self-interaction \citep{BernalGuzman2006b,GonzalezGuzman:2011,Schwabe:2016,Schive:2014,Schive:2014hza}, and this is an assumption we use in this paper as well.

The generalization of the GPP system to multi-states is as follows. Considering a  system described by $n$ states, and using scaled quantities such that we remove the physical constants (see e.g. \citep{GuzmanRivera2010,GuzmanLora2015}), the evolution of the system is described by the following GPP system of equations

\begin{eqnarray}
i\partial_t \Psi_1 &=& H\psi_1\nonumber \\
&&...\nonumber \\
i\partial_t \Psi_n &=&H\Psi_n \nonumber \\
\nabla^2 V &=& |\Psi_1|^2+...+|\Psi_n|^2,
\label{eq:gpp_general}
\end{eqnarray}

\noindent where $\Psi_1(t,{\bf x}),...,\Psi_n(t,{\bf x})$ represent the wave functions corresponding to the ground state of node-less wave functions until the most excited state with $n-1$ nodes and the Hamiltonian operator is $H = -\frac{1}{2}\nabla^2 + V$. Notice that the different states are evolved with separate Schr\"odinger equations, which are coupled through the gravitational potential $V$, in turn sourced by the addition of the density of each of the states. The first $n$ equations are evolution equations for each of the wave functions, whereas Poisson equation acts as a constraint of the evolution, since in general the gravitational potential $V$ will change in time as the density of each state $i$, specifically $\rho_i = |\Psi_i|^2$ is expected to change in general. This system of equations describes the evolution of the configurations in our analysis. The evolution of a general system then consists in the solution of the system of equations  (\ref{eq:gpp_general}) for consistent initial conditions. This approach defines an initial value problem for $n$ wave functions $\Psi_1,...,\Psi_n$ on the domain ${\bf x}\in R^3$ and $t\ge 0$, with the consistency condition that $V$ is the solution of Poisson equation in (\ref{eq:gpp_general}) at initial and further time. The initial conditions are free for the wave functions and $V$ is restricted by the densities $\rho_i$ of the various wave functions. 

We solve this problem numerically in 3D with axial symmetry following the recipe presented in \citep{BernalGuzman2006b,GonzalezGuzman:2011} and described as follows. Since the system of a head-on encounter is axysymmetric we define the problem on a two dimensional numerical domain in  cylindrical coordinates $r\in[0,r_{max}] \times z\in[z_{min},z_{max}]$. We discretize this domain with uniform resolution $r_i=i \Delta r $, $z_j=z_{min}+j\Delta z$, $i,j \in {\bf N}$, where we choose the spatial resolutions $\Delta r=\Delta z$. We set initial wave functions $\Psi_1(0,{\bf x}),...,\Psi_n(0,{\bf x})$ at each grid point of the discrete domain. Then we solve Poisson equation for $V$ using boundary conditions up to the quadrupole moment with a Successive Over Relaxation (SOR) method with optimal acceleration parameter. This sets the initial conditions to be evolved. The boundary conditions provided for the solution of Poisson equation is multipolar with leading monopolar term given by $V(r_{boundary})=-N/r_{boundary}$, where $N=\sum {N_i}$ is the total number of particles of all the states as defined below in (\ref{eq:W}) and $r_{boundary}$ is the distance of a point belonging to the boundary to the center of the domain.

The evolution over a discrete time step $\Delta t$ of the wave functions is carried out using the method of lines with a third order Runge-Kutta integrator. At each intermediate step of the integration of the wave functions we solve Poisson equation, which is mandatory since $V$ sources Schr\"odinger's equations. We do it using the same method as for the initial time. Then this procedure is repeated over a given number of time steps with constant $\Delta t$ until required. 

On the other hand, particularly involved are well posed boundary conditions for the Schr\"odinger type of equations. In practice what is best and has shown to work, is the implementation of a sponge. A sponge is implemented by  adding an imaginary potential near the boundary of the numerical domain $V\rightarrow V+V_{im}$. The effect of this imaginary potential is that produces a sink of particles that in the continuity equation reflects as $\partial \rho /\partial t +\frac{1}{2}\nabla \cdot [\Psi \nabla \Psi^{*} - \Psi^{*} \nabla \Psi]=2 V_{im}|\Psi|^2$. We thus implement this sponge outside a sphere of radius $r_{sponge}$ in a region outside of the configurations that are to be collided. In order to absorb as many modes as possible, we choose $V_{im}$ to have a smooth profile of the form. $V_{im}=-\frac{V_0}{2}[2+\tanh (r-r_{sponge})/\delta - \tanh(r_{sponge}/\delta)]$, where $\delta$ is the width of a transition region between the physical zone where $V_{im}=0$ and the zone of the sponge (see e.g. \citep{GuzmanRivera2010}).

% -----------------    Subsection
\subsection{Multi-state stationary configurations}

Since we will collide stationary configurations we provide a description on how they are constructed. For this we follow the recipe in \citep{UrenaBernal:2010}. First it is assumed that the wave functions in  system (\ref{eq:gpp_general}) to have spherical symmetry and  harmonic time dependence $\Psi_n(t,{\bf x}) = \psi_n(R)e^{i\omega_n t}$, where $R$ is the spherical coordinate. This particular time dependence implies the densities $\rho_n$ are time-independent and the spherical symmetry makes Poisson equation an ordinary equation along $R$, which makes easy to integrate the gravitational potential $V(R)$.

By imposing boundary conditions of regularity at the origin for the wave functions $\psi_n(R)$ and the gravitational potential $V(R)$, and isolation conditions, namely demanding $\psi_n(R)$ and its derivative to be zero, and a monopolar boundary condition for $V(R)$ at a numerical boundary far from the origin $R=R_{boundary}$, one sets a Strum-Liouville problem for the wave functions and the potential at the same time, and whose eigenvalues are the frequencies $\omega_n$. This problem is solved using a multi-shooting routine that pushes $R_{boundary}$ as far as possible, so that the boundary conditions are fulfilled \citep{UrenaBernal:2010}.

\begin{figure}
\centering
\includegraphics[width= 8cm]{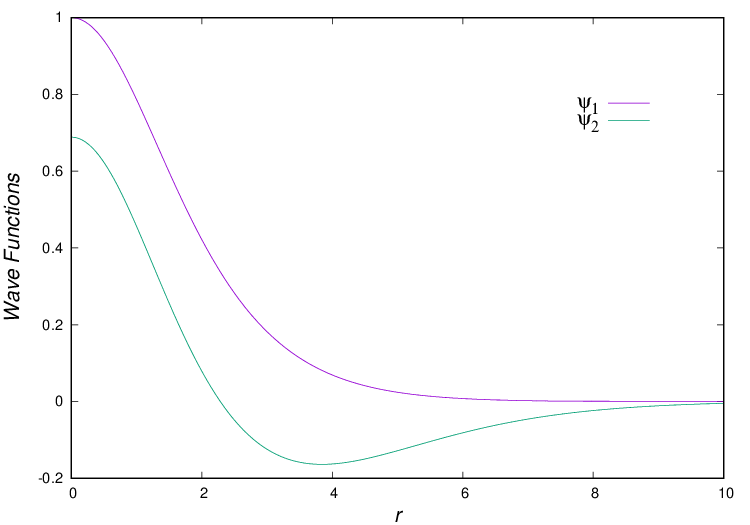}
\caption{Wave functions of the equilibrium configuration with two states. Notice that $\psi_1(r)$ has no nodes whereas $\psi_1(r)$ has one node. Should the Schr\"odinger equation be microscopic instead of macroscopic, the second state would be called the first excited state.}
\label{fig:equimulti}
\end{figure}

The results are the equilibrium wave functions $\psi_n(R)$ and the gravitational potential $V(R)$. In this paper we only consider at most three states and as an example we show in  Fig. \ref{fig:equimulti} the solution for a two state configuration. 
Notice that these configurations are constructed using spherical coordinates with resolution higher than that of the axisymmetric  numerical domain where we perform the collision. What we do with the equilibrium configuration is that we interpolate the wave functions into the axisymmetric domain as described below, before initiating the evolutions.

% -----------------    Subsection
\subsection{Setting initial data for two configurations that will collide}
\label{subsec:description}

Each of the configurations that will collide is an equilibrium configuration with a given number of states and a given set of wave functions. The head-on collision will happen along the $z-$axis starting at the same distance from the numerical origin at $(r=0,z=0)$. The evolution assumes the superposition of the wave functions of each of the two configurations that we use $\Psi_L$ for the configuration located initially at the left ($z<0$) and $\Psi_R$ for the configuration at the right ($z>0$). If $\Psi_L$ has $n$ states, we label its wave functions   $\psi_{L1},...,\psi_{Ln}$, if the configuration $\Psi_R$ has $m$ states, we label its wave functions  $\psi_{R1},...,\psi_{Rm}$. Assuming $1<m<n$, the wave function representing the superposition of the ground state configurations will be $\Psi_1=\psi_{L1}+\psi_{R1}$, the wave function for the first excited state will be $\Psi_2=\psi_{L2}+\psi_{R2}$ and so on and so forth until $\Psi_{m}=\psi_{Lm}+\psi_{Rm},...,\Psi_n = \psi_{Ln}$.  These superposed wave functions are those to be evolved using the GPP system (\ref{eq:gpp_general}).
In this paper the maximum value of $m$ or $n$ is 3. For the particular case of only ground state configurations, a single  wave function suffices to describe the collision of two configurations \citep{Choi2002,BernalGuzman2006b,ParedesMichinel2016,Schwabe:2016}, as well as for the study of structure formation \cite{Schive:2014}. This corresponds to the case $m=n=1$ and the unique wave function is the superposition of the wave function of each configuration $\Psi_1=\psi_{L1}+\psi_{R1}$, and the evolution is ruled by only one Schr\"odinger equation in (\ref{eq:gpp_general}). 
In order to put this idea into practice one needs to interpolate the multi-state equilibrium configurations constructed in the previous section and center them as follows. We center $\Psi_L$ at the point $(r,z)=(0,-z_0)$ and $\Psi_R$ at $(r,z)=(0,z_0)$. In order to set the configurations in the axial domain here the evolution is to be carried out, we interpolate the numerical solution of the wave functions of $\Psi_L$ and $\Psi_R$. We choose $z_0$ so that the two configurations are far enough from one another in order for the interference term $\langle\psi_L,\psi_R\rangle$ to be smaller than round-off error. 
The collision will be head-on and along the $z$ axis, thus the addition of linear momentum to the initial blobs is applied through the transformation of the wave functions as follows $\psi_L\rightarrow e^{i{\bf p_z \cdot x}}\psi_L$ and $\psi_R \rightarrow e^{-i{\bf p_z \cdot x}}\psi_R$.
% --------------     Subsection
%
\subsection*{Diagnostics}
Assuming the wave functions to be orthogonal facilitate the estimates of some physical quantities. The expectation value of an operator $\hat{A}$ would be given by
\begin{eqnarray}
 A=\sum_{k} \int \Psi_{k}^{*} \hat{A} \Psi_{k} d^3x .
\end{eqnarray}

\noindent where the summation runs over the number of wave functions involved. Of particular importance are the kinetic energy, the gravitational energy and the number of particles associated to each wave function, that we calculate respectively as 

\begin{eqnarray}\label{eq:K}
K_i&=&-\frac{1}{2}\int \Psi^{*}_{i} \nabla^2 \Psi_{i} d^3 x,\nonumber\\
W_i&=&\frac{1}{2}\int \Psi^{*}_{i} V \Psi_{i} d^3 x,\nonumber\\
N_i&=&\int \rho_i d^3 x,\label{eq:W}
\end{eqnarray}

\noindent for each wave function $\Psi_i$. The integration is carried out over the whole numerical domain. As a particular case there is the collision of two configurations, like in \citep{Schive:2014,Schive:2014hza,Schwabe:2016,BernalGuzman2006a,BernalGuzman2006b}, where only one wave function for the two configurations was prescribed.

One of the exciting properties of binary GPP configurations is that they are able to form bounded final systems depending on the value of the total energy given by $E=K+W$. Whenever $E<0$ it happens that the gravitational binding is sufficiently strong as to confine the individual configurations in a bounded region whilst a common gravitational potential well arises. Otherwise, when $E>0$, GPP solutions are dispersive because the gravitational binding is strong enough to keep individual configurations confined in an bounded region and therefore the final configuration tears apart into out-going states. That the profile of the resulting states is pretty similar to the initial one, justifies the configuration to be some times called solitonic.

% -------------------------------------------
% ----->     Section     <-----
% -------------------------------------------
\section{Tests}
\label{sec:tests}

Before we move to the central study of the paper, we show that the code passes two basic tests. We verify that some well known cases are properly reproduced, namely, the evolution of a single configuration laying in an excited state and the collision of two configurations in the ground state.

% --------------     Subsection
\subsection{Tests with single configurations}

A first test consists in reproducing the evolution of  equilibrium configurations in different multi-states. The density profile of each configuration must remain stationary while the wave function associated to each state oscillates. 

According to our notation above, for a configuration in the ground state case, a single wave function $\{\Psi_1\}$ suffices to describe the evolution of the system, whereas for the case with two states we use two $\{\Psi_1,\Psi_2\}$ and three functions $\{\Psi_1,\Psi_2,\Psi_3\}$ for a configuration with three states. 

\begin{figure}
\centering
\includegraphics[width= 4.25cm]{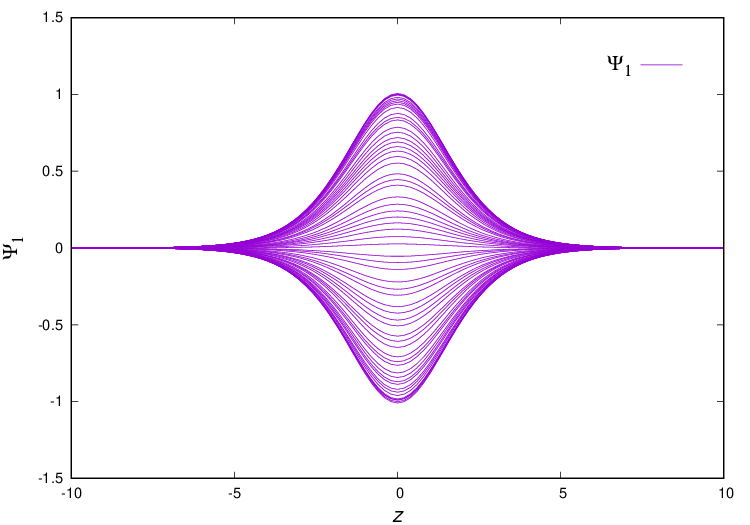}
\includegraphics[width= 4.25cm]{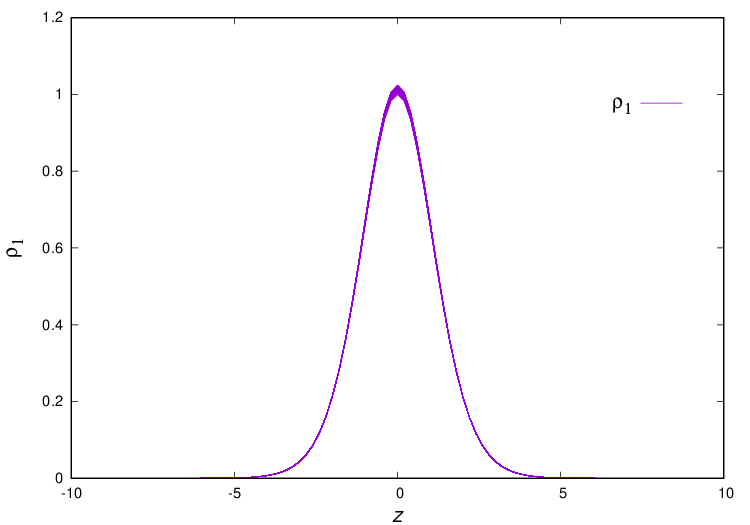}
\includegraphics[width= 4.25cm]{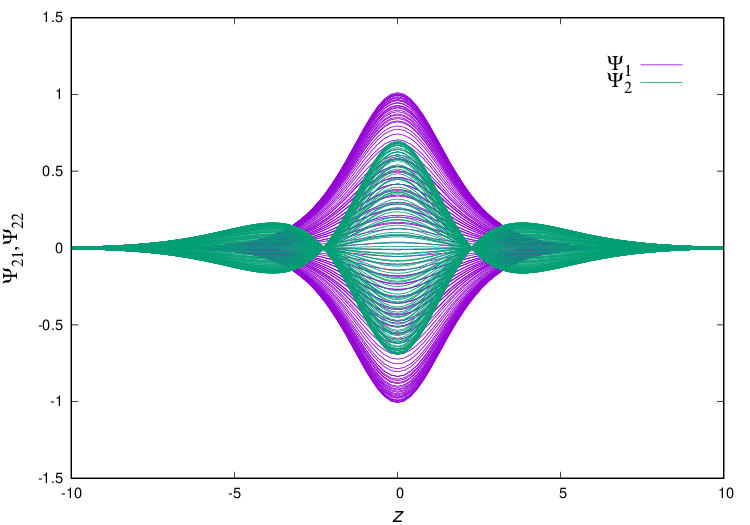}
\includegraphics[width= 4.25cm]{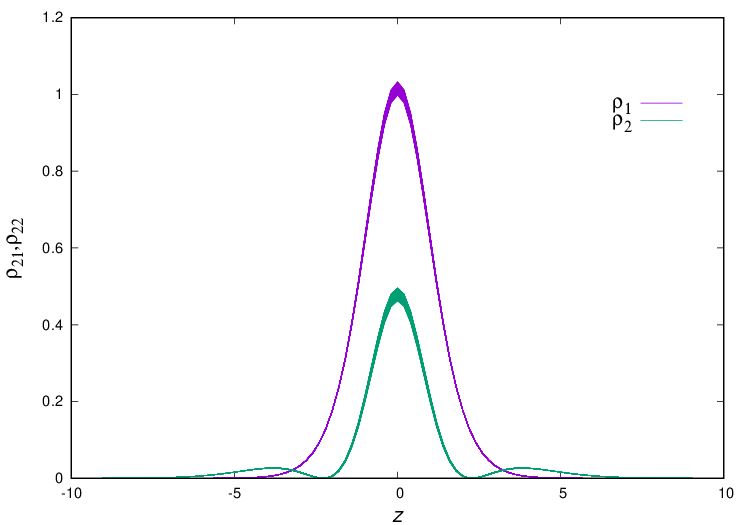}
\includegraphics[width= 4.25cm]{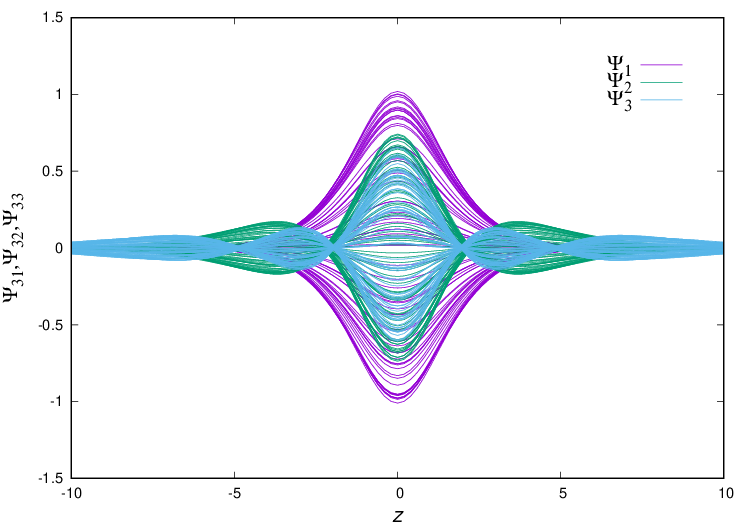}
\includegraphics[width= 4.25cm]{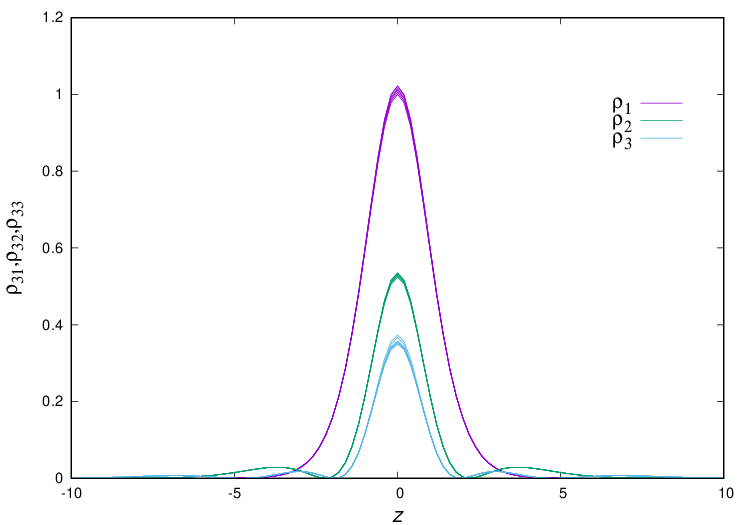}
\caption{Evolution of the wave functions and densities of configurations 1, 2 and 3. The density of each state remains nearly stationary whereas the wave function is oscillating. The tests run until $t=100$ in code units. The evolution of these configurations was calculated with the axially symmetric code on the domain $r\in [0,20],~ z\in[-20,20]$ with resolution $\Delta r= \Delta z = 0.2$ and the configuration set at the coordinate center.}
\label{fig:single_evolution}
\end{figure}

In Fig. \ref{fig:single_evolution} we show snapshots of the wave functions for each of the three cases and the corresponding densities defined as $\rho_i=|\Psi_i|^2$. The plots show that each wave function oscillates whereas the corresponding density remains nearly stationary with small amplitude oscillations. This behavior indicates that the code solves the system of equations (\ref{eq:gpp_general}) consistently with the theory and with previous numerical results \citep{UrenaBernal:2010}.

% --------------     Subsection
\subsection{\textbf{Second test}: collision of two ground state configurations}
\label{subsec:single}

As an extra test for our code we choose the collision of two ground state configurations. For this we use similar parameters as those used in \cite{BernalGuzman2006b,GonzalezGuzman:2011}. This case requires only one wave function $\Psi_1=\Psi_{L1}+\Psi_{R1}$ according to the notation described above. In order to add  momentum along the head-on direction, we redefine the wave function of both configurations as follows

\begin{eqnarray}\nonumber
\Psi_{L1} &\rightarrow& e^{i p_z \cdot z} \Psi_{L1},\\
\Psi_{R1} &\rightarrow& e^{ -i p_z \cdot z}\Psi_{R1},\nonumber
\end{eqnarray}

\noindent with $\Psi_{1}=\Psi_{L1}+\Psi_{R1}$ as the initial wave function. One then solves Poisson equation at initial time and continues the evolution using the system (\ref{eq:gpp_general}) for $n=1$.

\begin{figure}
\centering
\includegraphics[width= 4.25cm]{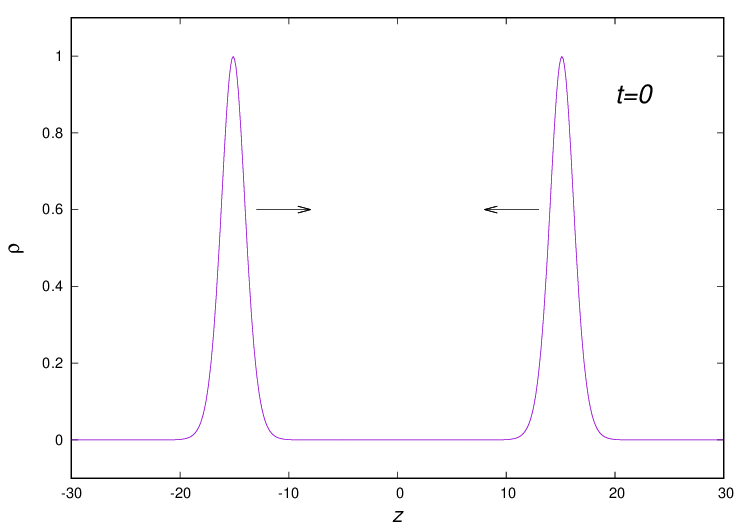}
\includegraphics[width= 4.25cm]{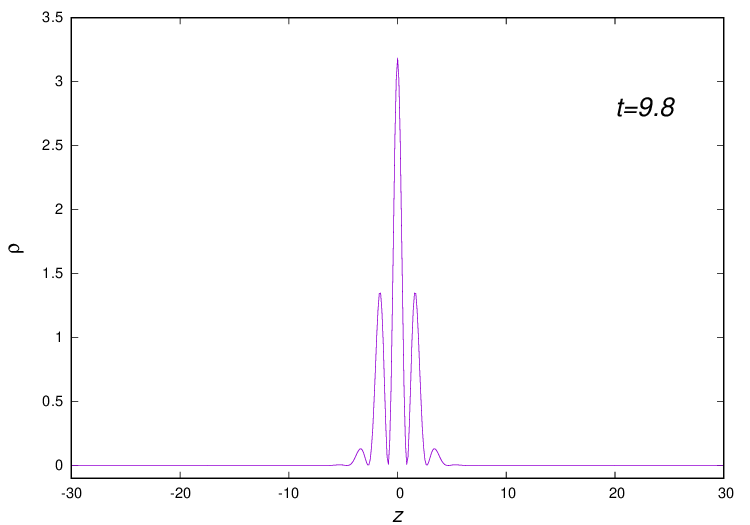}
\includegraphics[width= 4.25cm]{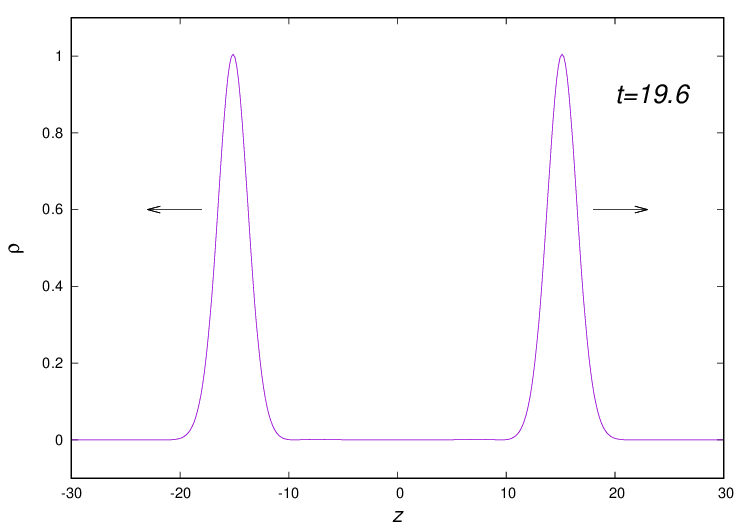}
\includegraphics[width= 4.25cm]{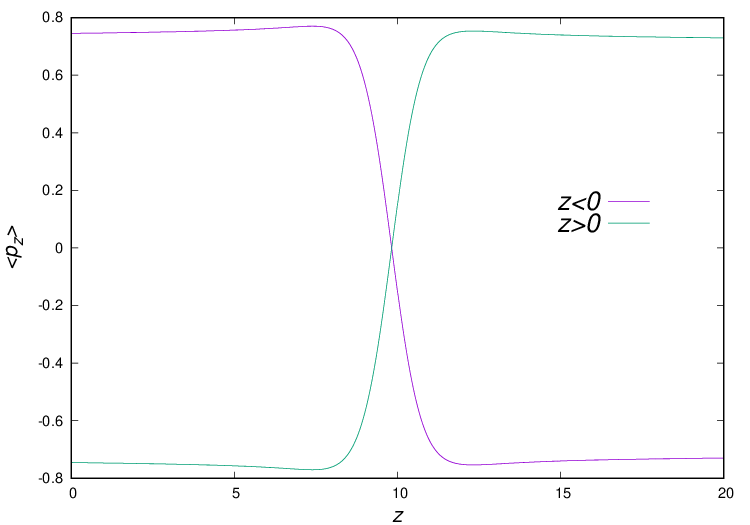}
\caption{Solitonic case of a ground state - ground state collision. We show the density $\rho_1$ at various times along the $z$-axis, before, during and after the interaction. The arrows indicate the direction of motion of the two configurations. We also show the transfer of linear momentum integrated in the left and right domains. For this we show the expectation value of the operator $p_z$ in the half domain $z<0$ and in the other half domain $z>0$. The evolution was calculated on the numerical domain $r\in [0,30],~ z\in[-30,30]$ with resolution $\Delta r= \Delta z = 0.12$ and the configurations were initially located using  $z_0=15$.}
\label{fig:case1_1_solitonic}
\end{figure}

{\it High initial velocity case.} In order to obtain the solitonic behavior we  use $p_z=1.5$ in code units, which corresponds to a large initial linear momentum giving rise to a positive total energy 
 $E=K+W>0$ and therefore the system turns out to be unbounded.
In Fig. \ref{fig:case1_1_solitonic} we show the density profiles at initial time and various snapshots during and after the interaction between the two configurations. We also show a  plot of the linear momentum transfer from the half domain $z<0$ to the half domain $z>0$ during the process and in the opposite direction. For this we calculate the expectation value of the linear momentum along the head-on direction ($z$), in the two regions using $\langle p_z \rangle = \int_{D} \Psi^{*}_{1}(-i \frac{\partial}{\partial z})\Psi_{1} d^3 x$, where the domain $D$ can be  the half domain $z<0$ or the other $z>0$. The plot shows that the momentum transfered from the left domain $z<0$ where initially the $L$ configuration (with positive momentum) is set, to the right domain $z>0$ and from the right domain (with the $R$ configuration initially with negative momentum) to the left, shows the same behavior. An important observation is that the density profiles at time $t=0$ and $t=19.6$ is not exactly the same, a comparison would reveal small differences in shape, at first glance final individual configurations are wider than the initial ones. Strictly speaking, for solitons the density is the same in the asymptotic past and future times \cite{Rajaraman:1982is}. In the present study, equilibrium configurations are not solitons due to the presence of  the gravitational potential involved, which influences the shape in the asymptotic time, aside of the fact that the simulation is being carried out in a finite spatial and temporal domain, which prevents a strict asymptotic analysis.

\begin{figure}
\centering
\includegraphics[width= 4.25cm]{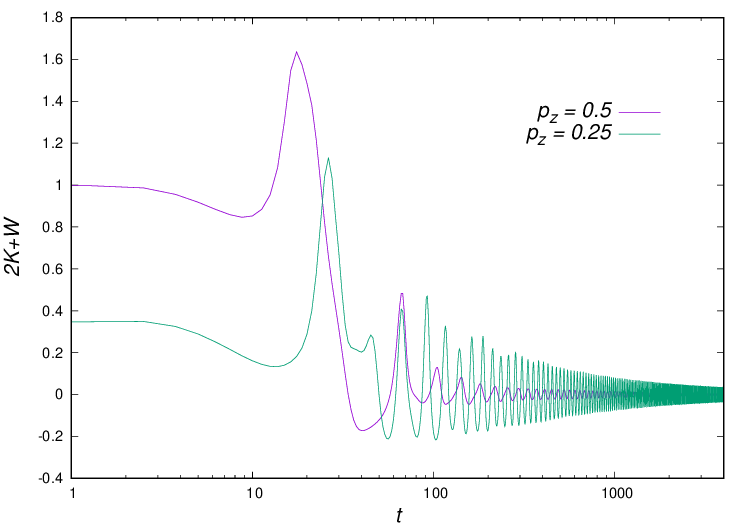}
\includegraphics[width= 4.25cm]{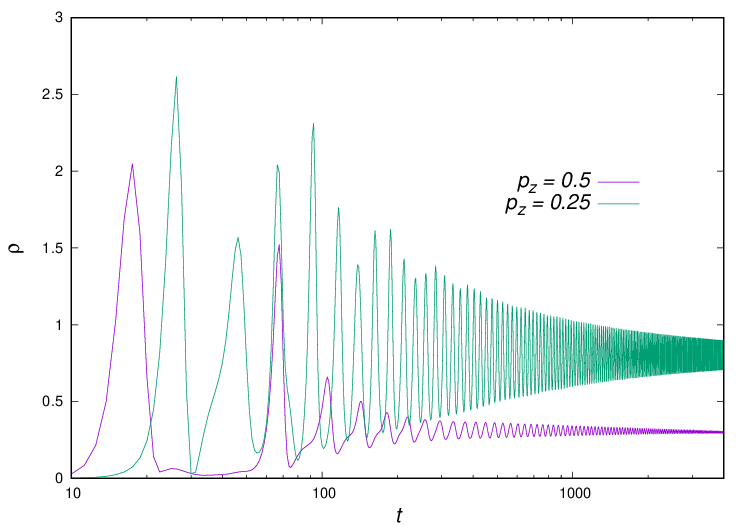}
\includegraphics[width= 4.25cm]{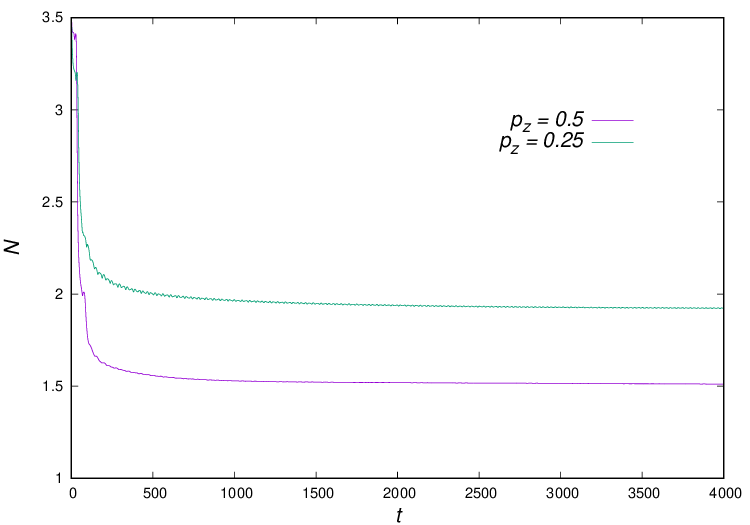}
\includegraphics[width= 4.25cm]{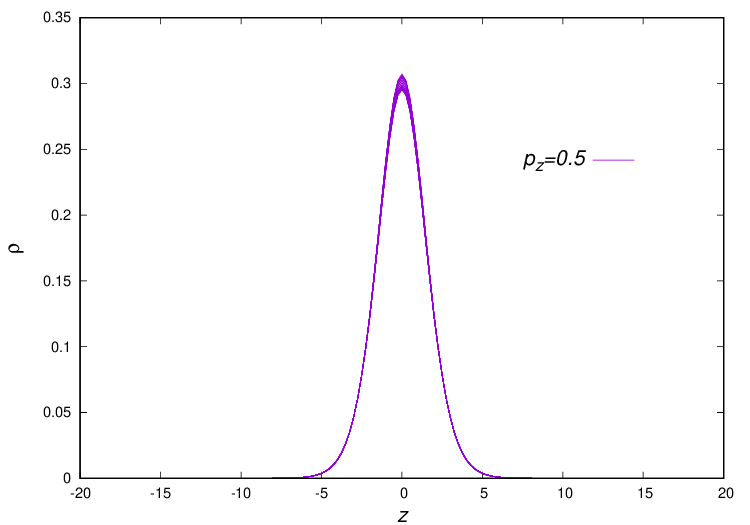}
\caption{Merger of ground state - ground state case. We show diagnostics of configurations that merge, for initial momentum $p_z=0.5,~0.25$. First the quantity $2K+W$, that approaches asymtotically to zero in time. Second, the value of the density in the center of the merger, showing an oscillatory behavior with decreasing amplitude. We also calculate $N_1=\int \rho_1 d^3 x$, which is the total number of particles within the whole numerical domain, indicating that part of the material gets off the domain and during the relaxation process remains nearly constant. Finally we show a sample of snapshots of the density of the final configuration for the case $p_z=0.5$, that shows an oscillatory behavior.
The evolution was calculated on the domain $r\in [0,30],~ z\in[-30,30]$ with resolution $\Delta r= \Delta z = 0.12$ and the configurations were initially located with  $z_0=15$.}
\label{fig:binary11merger}
\end{figure}

{\it Low velocity case.} The scenario for the merger of the two configurations requires the momentum to be smaller. For this we choose two values of the head-on momentum $p_z=0.25,~0.5$ that help showing the behavior of the system and the relaxation process that leads to a final structure in a long term evolution. A helpful quantity that determines whether a system is approaching equilibrium is $2K+W$, which in theory is zero for a virialized system, and has been used in the past to monitor the Gravitational Cooling process of configurations ruled by the GPP system \citep{Seidel-Suen:1990,GuzmanUrena2006} . Based on this diagnostics, we show in Fig. \ref{fig:binary11merger} the behavior in time of this quantity, together with the central value of the density $\rho_1=|\Psi_{1}|^2$. At the beginning $2K+W$ is positive
and far from zero because the head-on momentum involved with the kinetic energy of the system dominates; after the two configurations have merged, this quantity oscillates with  a decreasing amplitude and is expected to be zero asymptotically, which would indicate the system tends towards a virialized final state. 

The central value of the density on the other hand also shows an asymptotic behavior, that is, it approaches a constant value after oscillating with a decreasing amplitude. This is an indication that the configuration resulting from the merger tends to be time-independent asymptotically. Notice that the two values of the head-on momentum ($p_z=0.25,0.5$) result in two different central densities of the final configuration. The reason is that even though the system is bounded, part of the density of the initially moving configurations escapes from the gravitational potential. We use two values of $p_z$ in order to illustrate that the faster the motion of the configurations, the smaller the final configuration. This is illustrated in the bottom panel of Fig. \ref{fig:binary11merger}, where we show $N_1$ in time. The fact that this quantity decreases initially indicates that a finite amount of particles of the system  gets off the domain, and also the fact that it approaches a constant value indicates that a finite number of particles remains trapped in the gravitational potential.
Finally in this Figure we also show snapshots of the density $\rho_1$ for the case $p_z=0.5$, during the final stages of the simulation between $t=3000$ and $4000$ in code units. The density oscillates as shown by the plot of its central density, with a decreasing amplitude. 

In theory an average in time of this density could serve to fit the profile of the final configuration as done for instance in 3D simulations \cite{Schive:2014,Schwabe:2016}. Indeed, as a part of our tests, here we make such a procedure in order to obtain a final density profile for the final configuration of the merger and be able to compare with profiles from 3D structure formation simulations. By carrying out a time-average of the radial density profile of the merger along a period of oscillation, we compute a mean density profile which is  illustrated later in Fig. \ref{fig:case22merger_slope}. In previous studies, specifically based on the profile of structures after mergers \cite{Schwabe:2016}, the final density profile splits into a soliton-like core and power-law tail, such layout is due to the fact that the resulting configuration has non-zero angular momentum. In contrast, for the head-on case, we find that the final configuration is made of a pure solitonic state whose average density profile is perfectly described by a Gaussian fit given by:
%rho0= 0.660636271967
%x0= 0.506427504546
%sigma= 1.36220941154
\begin{eqnarray}
\frac{\rho(r)}{\rho_0}&=&e^{-\left(\frac{x}{6.37}\right)^2},
%\ln \rho_{core}^{(g)}&=& -0.03462282 z^2  -0.90574324z -0.31030531, \\
%\nonumber \ln \rho_{core}^{(c)}&=&-0.102(\ln z)^3 -0.3501(\ln z)^2\\ &&-0.4481(\ln z) -0.236\\
 %log rho = -0.03462282*z**2--0.90574324*z -0.31030531
\label{eq:fit_GG}
\end{eqnarray}

\noindent which is a behavior consistent with the attractor properties of ground state configurations. Fits for the final core resulting from 3D simulations \citep{Schive:2014,Schive:2014hza} is reported to obey $\rho\sim(1+0.091(r/r_c))^{-8}$. In contrast, our final distribution obeys \eqref{eq:fit_GG} which turns out to be much more shallow than the former. On the other side, while our final profile turns out to be purely solitonic over the whole domain, the fit from \citep{Schive:2014} decays more slowly at the edge following a NFW envelope. Such an apparent discrepancy is expected to happen due to the different dynamics with structure formation in \citep{Schive:2014}, binary mergers with orbital momentum in \citep{Schwabe:2016} and our binary head-on case.  Tangential encounters involve interference mechanisms and tidal effects that do not seem to match the soliton profile of the final state at some point in the edge of the core giving rise to a core-tail profile. Nevertheless, we shall see along the next section that when multi-states are involved in the collision such a behavior of the tail of the final configuration arises.

% -------------------------------------------
% ----->     Section     <-----
% -------------------------------------------
\section{Results}
\label{sec:results}

% ----->.    Subsection
\subsection{Ground state vs a two-states collision}

In this case we set the state $\Psi_L$ to be the multi-state configuration characterized by two wave functions $\{\Psi_{L1},\Psi_{L2}\}$, whereas the state $\Psi_R$ is characterized only by a single wave function $\Psi_{R1}$. Recall these wave functions correspond to equilibrium spherically symmetric configurations, and in order to add linear momentum momentum in the head-on direction, these wave functions are redefined as follows

\begin{eqnarray}
\Psi_{L1} &=& e^{i p_z \cdot z} \Psi_{L1}\nonumber\\
\Psi_{L2} &=& e^{i p_z \cdot z}\Psi_{L2}\nonumber\\
\Psi_{R1} &=& e^{ -i p_z \cdot z}\Psi_{R1}\nonumber
\end{eqnarray}

\noindent In this case the system (\ref{eq:gpp_general}) requires the solution of two Schor\"odinger equations and the gravitational potential is sourced with two densities $\rho_1 = |\Psi_1|^2$ and $\rho_2=|\Psi_2|^2$, where $\Psi_1=\Psi_{L1}+\Psi_{R1}$ and $\Psi_2=\Psi_{L2}$.
We study two scenarios, one exploring a high linear momentum, that eventually could show the solitonic behavior found in the pure ground state configurations, and a second one of small momentum that allows the merger of the two configurations.

\begin{figure}
\centering
\includegraphics[width= 4.25cm]{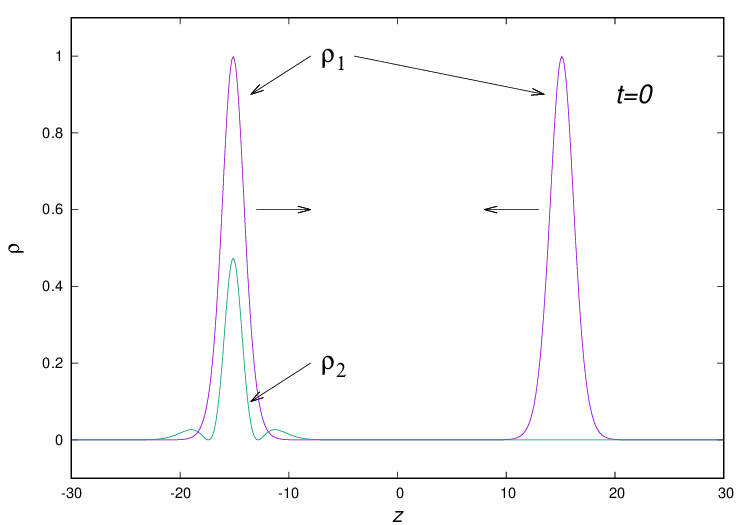}
\includegraphics[width= 4.25cm]{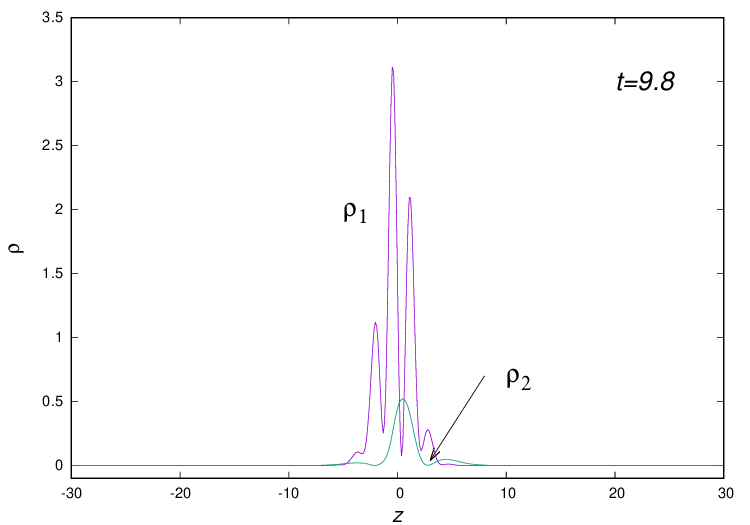}
\includegraphics[width= 4.25cm]{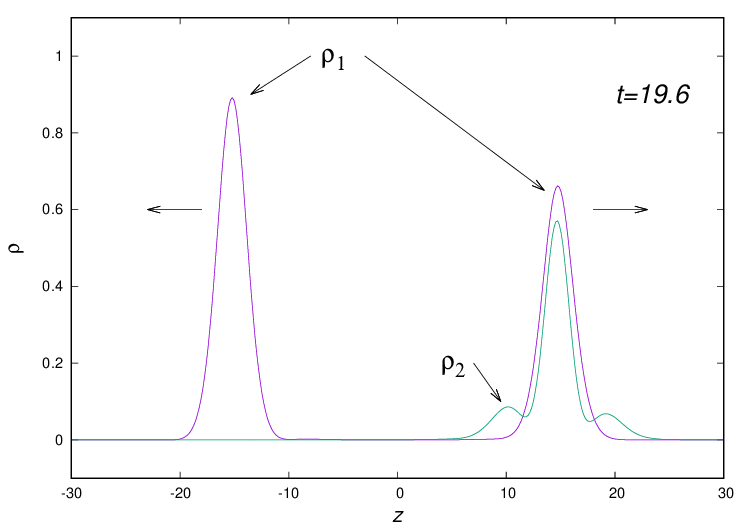}
\includegraphics[width= 4.25cm]{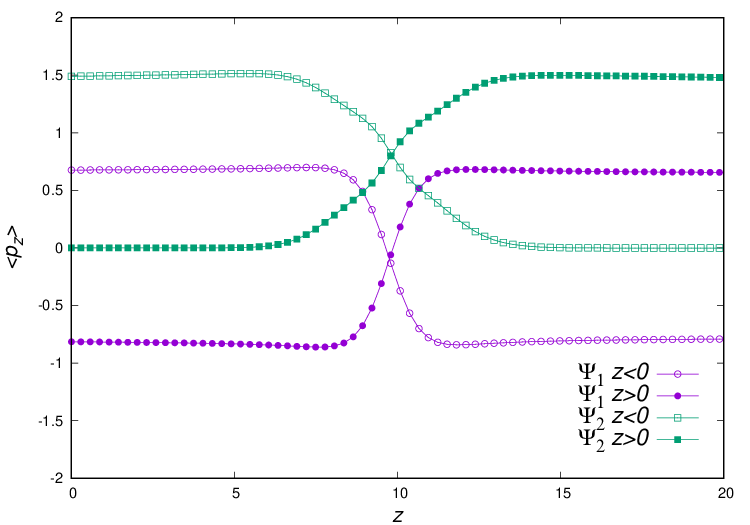}
\caption{Solitonic case of a ground state - two-states encounter. Density of the binary configuration at various times along the $z$-axis, before, during and after the interaction. The unlabeled arrows indicate the direction of motion of the two configurations. We also show the transfer of linear momentum integrated in the $z<0$ and $z>0$ half domains, for each of the two states that are being evolved.  For this we show the expectation value $\langle p_z \rangle$ integrated in the domain $z<0$ and in the domain $z>0$ for both, the ground state and the excited state wave functions. The evolution was calculated on the domain $r\in [0,30],~ z\in[-30,30]$ with resolution $\Delta r= \Delta z = 0.12$ and the configurations were initially located with  $z_0=15$.}
\label{fig:case2_1_solitonic}
\end{figure}

% ------------------------
{\it High velocity case}. For this we use the linear momentum  $p_z=1.5$, in order for the two configurations to go through each other. The result of the interaction between the ground state configuration and the configuration made of two first states is shown in Fig. \ref{fig:case2_1_solitonic}. Likewise in the previous case of ground state configurations, we show snapshots of the density and the momentum transfer during the interaction. 

During the stage at which individual configurations overlap inside a region around $z=0$, an interesting behavior can be noticed. As the  snapshot at $t=9.8$ shows, only $\rho_1$ shows an interference pattern with dominant amplitude and a while after the interaction, the amplitude of this density decreases. On the other hand $\rho_2$ does not produce such a pattern and its amplitude is enhanced. Besides, notice that for this multi-state, the initial and final profiles differ importantly, the initially excited state solution loses its nodes after the collision.

We measure the momentum transference by calculating $\langle p_z \rangle_{1,2}$ for the two wave functions. $\langle p_z\rangle_1$ is non-zero initially, positive in the left half-domain and negative in the right half-domain, and after the interaction it is transferred from one half-plane to the opposite one. On the other hand $\langle p_z\rangle_2$ is non-zero only for the configuration on the left, and after the interaction its value is transferred to the opposite half-domain.

\begin{figure}
\centering
\includegraphics[width= 4cm]{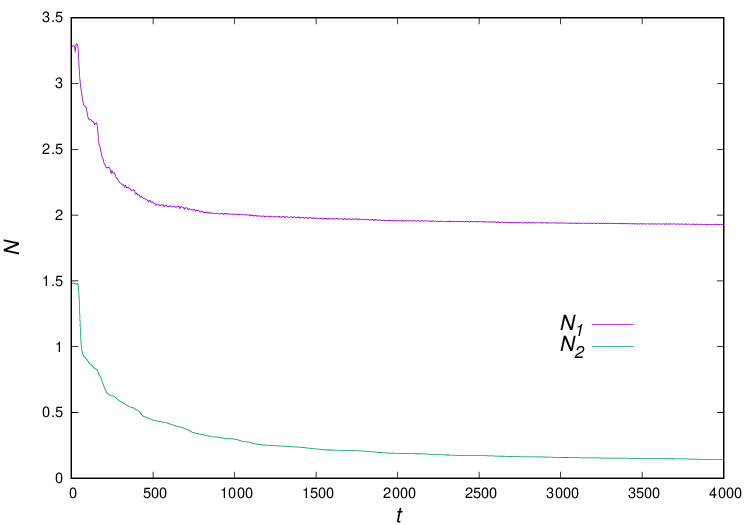}
\includegraphics[width= 4cm]{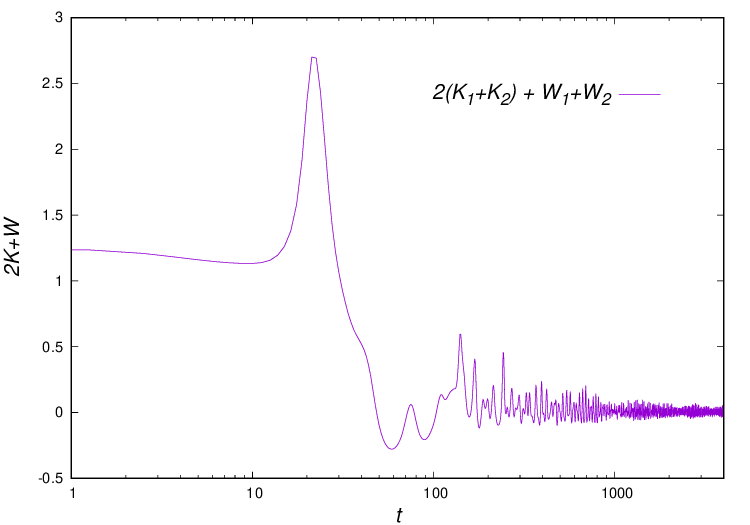}
\includegraphics[width= 4cm]{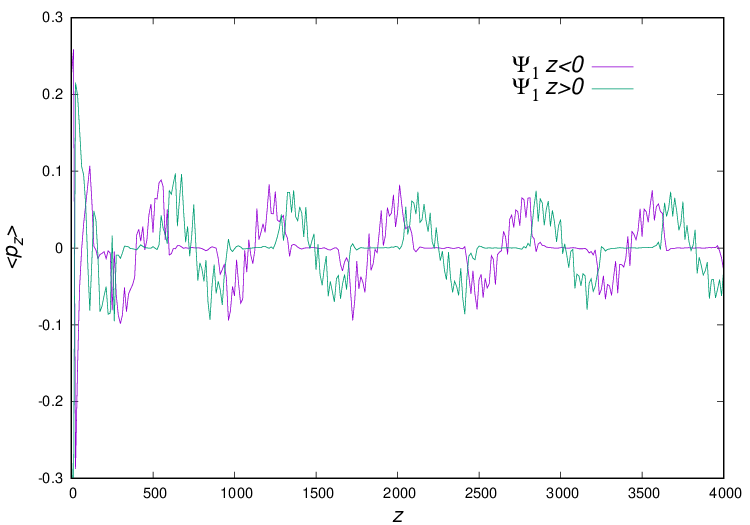}
\includegraphics[width= 4cm]{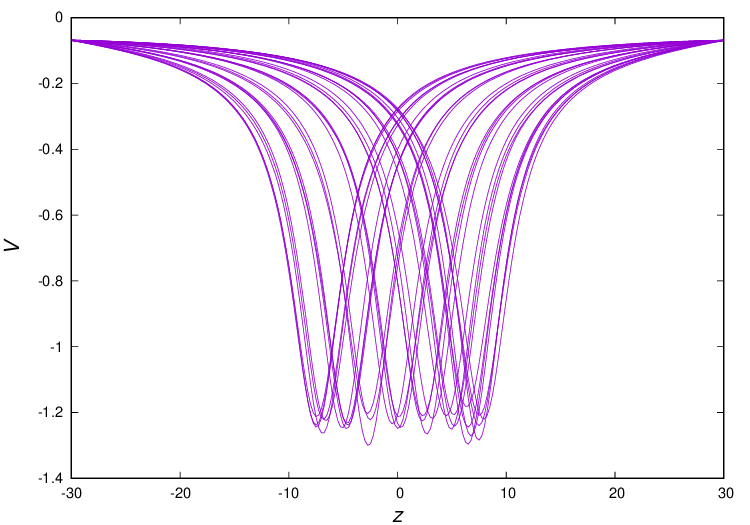}
\caption{Merger case of a ground state - two-states encounter. First we show $N_1$ and $N_2$ of the system, showing it approaches a constant number of particles during the process. Second we plot the quantity $2K_1 + 2K_2 + W_1 + W_2$, which oscillates with decreasing amplitude around zero. This indicates that the system tends toward a state of relaxation. In the bottom we show $\langle p_z \rangle$ integrated in the two halves of the domain for $\Psi_1$. The oscillatory behavior indicates that even though the system as a whole shows a relaxation process, the particles are being transferred from one half of the domain to the other. Finally we show snapshots of the gravitational potential $V$ along the $z$ axis for various times after the first merge. The position of the minimum oscillates from negative to positive values of $z$. The evolution was calculated on the domain $r\in [0,30],~ z\in[-30,30]$ with resolution $\Delta r= \Delta z = 0.12$ and the configurations were initially located with  $z_0=15$.}
\label{fig:case2_1_merger}
\end{figure}

% ------------------------
{\it Low velocity case.}  We produce a merger case scenario using $p_z=0.5$ which has negative total energy and thus corresponds to a bounded system. For this case, we show that the quantity $2K_1+2K_2 + W_1 +W_2$ in Fig. \ref{fig:case2_1_merger} oscillates and approaches to zero asymptotically in time. We also show the total energy and the integrals $N_1$ and $N_2$ illustrating how the system loses particles and then approaches a stationary regime. Nevertheless, since we launched the two configurations with the same head-on momentum, and the center of mass is off the center of coordinates, since there are no dissipative effects other than the gravitational cooling thorough the emission of particles, the center of mass of the system oscillates. This is illustrated by the value of $\langle p_z \rangle$ integrated in the two halves of the domain $z>0,~ z<0$, showing that the momentum is being transferred back and forth between the two regions. We include this scenario because it seems to show the properties required for dark matter in extreme scenarios like the bullet cluster, where, luminous matter seems to slow down with respect to the dark matter distribution. Finally we present snapshots of the  gravitational potential, whose minimum oscillates during various cycles without a noticeable dissipation.

% ----->.    Subsection
\subsection{Ground state vs a three-states configuration collision}

Like in the previous case, we set the state $\Psi_L$ to be the multi-state configuration characterized by three wave functions $\{\Psi_{L1},\Psi_{L2},\Psi_{L3}\}$, whereas the state $\Psi_R$ is characterized only by one wave function $\Psi_{R1}$. In order to add linear momentum  in the head-on direction, these wave functions are redefined as follows

\begin{eqnarray}
\Psi_{L1} &=& e^{i p_z \cdot z} \Psi_{L1}\nonumber\\
\Psi_{L2} &=& e^{i p_z \cdot z}\Psi_{L2}\nonumber\\
\Psi_{L3} &=& e^{i p_z \cdot z}\Psi_{L3}\nonumber\\
\Psi_{R1} &=& e^{ -i p_z \cdot z}\Psi_{R1}.\nonumber
\end{eqnarray}

\noindent In this case the system (\ref{eq:gpp_general}) requires the solution of three Schor\"odinger equations and the gravitational potential is sourced with three densities $\rho_1 = |\Psi_1|^2$, $\rho_2=|\Psi_2|^2$ and $\rho_3$, where $\Psi_1=\Psi_{L1}+\Psi_{R1}$, $\Psi_2=\Psi_{L2}$ and $\Psi_3=\Psi_{L3}$.

\begin{figure}
\centering
\includegraphics[width= 4.25cm]{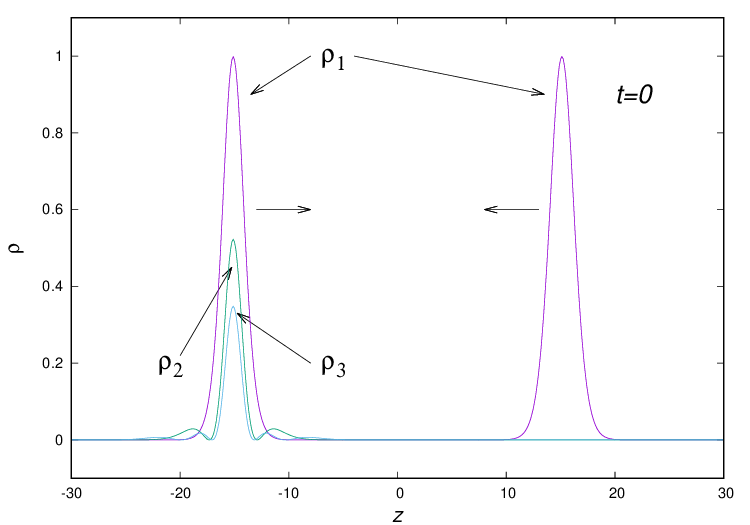}
\includegraphics[width= 4.25cm]{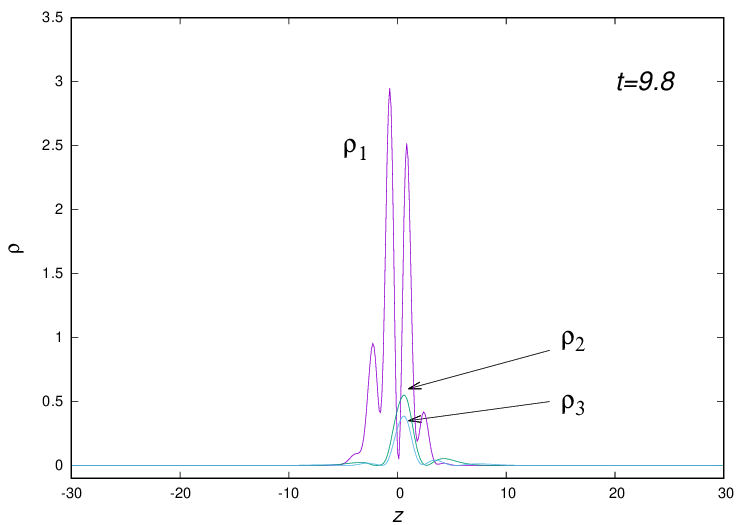}
\includegraphics[width= 4.25cm]{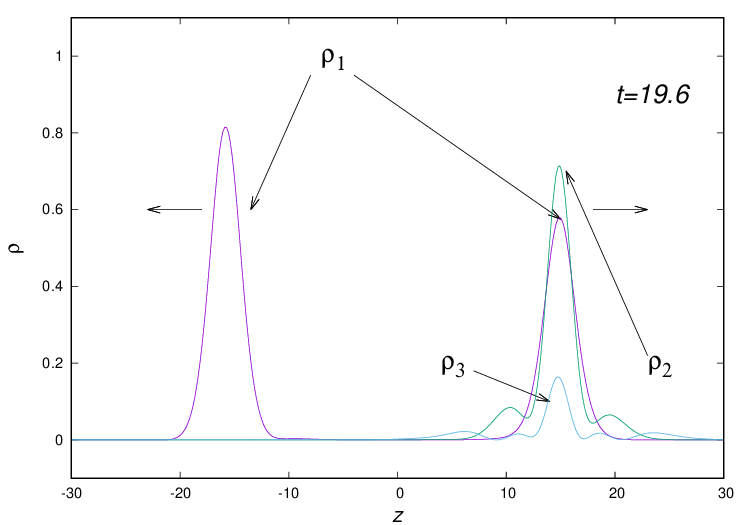}
\includegraphics[width= 4.25cm]{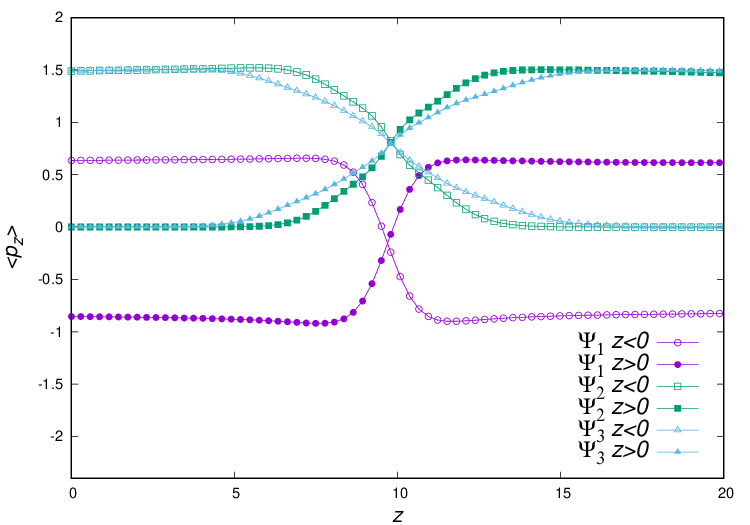}
\caption{Solitonic case of a ground state - three-states configuration encounter. Density of the binary configuration at various times along the $z$-axis, before, during and after the interaction. The arrows indicate the direction of motion of the two configurations. We also show the transfer of linear momentum integrated in the left and right domains for each of the two states that are being evolved.  For this we show $\langle p_z \rangle$ integrated in the domain $z<0$ and in the domain $z>0$ for both, the ground state and the excited state wave functions. The evolution was calculated on the domain $r\in [0,30],~ z\in[-30,30]$ with resolution $\Delta r= \Delta z = 0.12$ and the configurations were initially located with  $z_0=15$.}
\label{fig:case3_1_solitonic}
\end{figure}

% ------------------------
For a {\it high velocity case} we again use the momentum $p_z=1.5$ and show the densities of each state in Fig. \ref{fig:case3_1_solitonic}. The results are similar to those in the previous case. The density $\rho_1$ shows an interference pattern whereas the other tow states approximately retain their profile during the evolution. After the interaction it happens that the first excited state is amplified, whereas the second excited state flattens. The solitonic behavior is not perfect again and the densities associated to each state suffer deformations due to the interaction between the two configurations through the gravitational potential.

% ------------------------
{\it Low velocity case.} By applying the same head-on momentum as in the previous cases $p_z=0.5$, the results are pretty similar to those of the merger of a ground state vs a two-states configuration. In brief, the final configuration tends globally toward a virialized state, however the final blob oscillates in space around the center of coordinates.

% ----->.    Subsection
\subsection{Two-states vs two-states configurations collision}

Now we show the result of the collision of two non-ground state configurations. For this we choose the two configurations to have two states. In this case the configuration at the left has two wave functions $\{\Psi_{L1},\Psi_{L2}\}$  and the configuration at the right has also two wave functions $\{\Psi_{R1},\Psi_{R2}\}$.

The addition of linear momentum reassigns the wave functions in the following manner

\begin{eqnarray}
\Psi_{L1} &=& e^{i p_z \cdot z} \Psi_{L1},\nonumber\\
\Psi_{L2} &=& e^{i p_z \cdot z} \Psi_{L2},\nonumber\\
\Psi_{R1} &=& e^{ -i p_z \cdot z}\Psi_{R1},\nonumber\\
\Psi_{R2} &=& e^{ -i p_z \cdot z}\Psi_{R2}.\nonumber
\end{eqnarray}

\noindent Finally, following the description in subsection \ref{subsec:description}, the wave function of the ground state is given by  $\Psi_1=\Psi_{L1}+\Psi_{R1}$, whereas the first excited state wave function of the system is $\Psi_2=\Psi_{L2}+\Psi_{R2}$.

In this case the system (\ref{eq:gpp_general}) requires the solution of two Schor\"odinger equations and the gravitational potential is sourced with two densities $\rho_1 = |\Psi_1|^2$ and $\rho_2=|\Psi_2|^2$. Like in the previous cases, in order to learn about the general behavior of the system we execute high velocity and low velocity cases.

\begin{figure}
\centering
\includegraphics[width= 4.25cm]{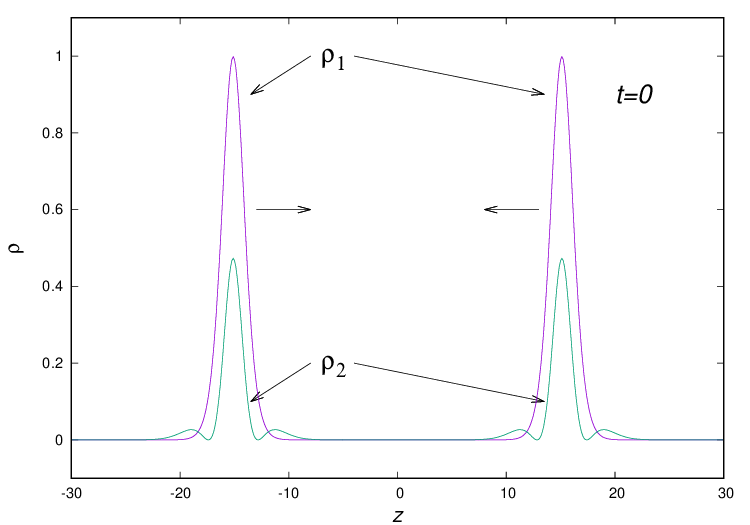}
\includegraphics[width= 4.25cm]{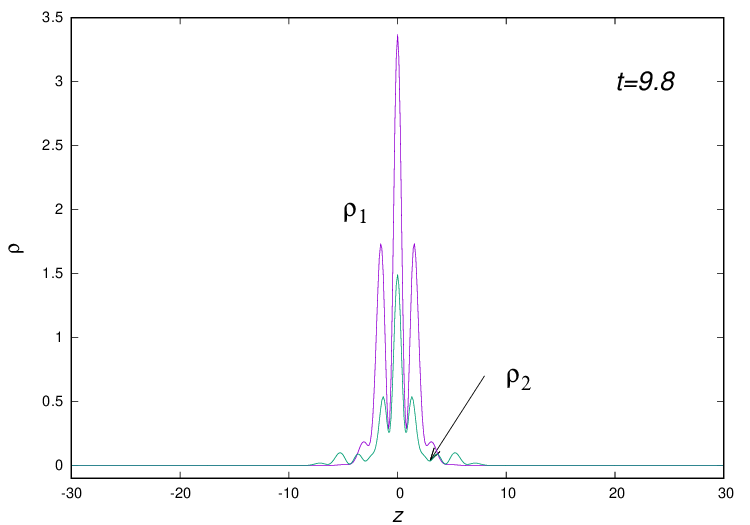}
\includegraphics[width= 4.25cm]{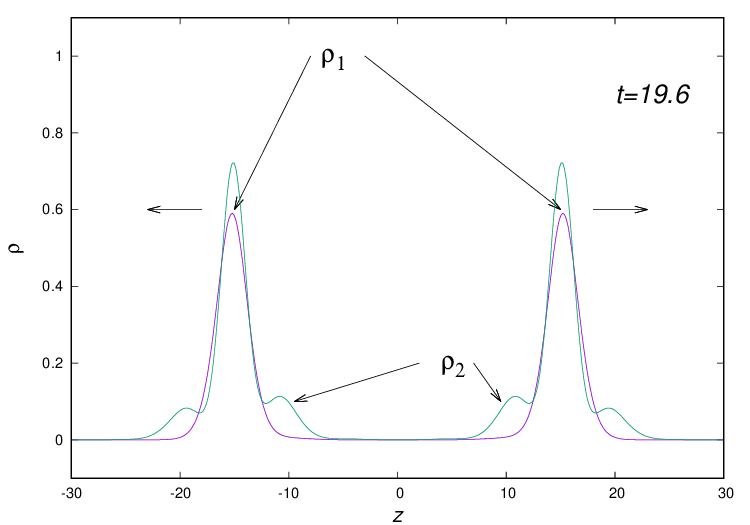}
\includegraphics[width= 4.25cm]{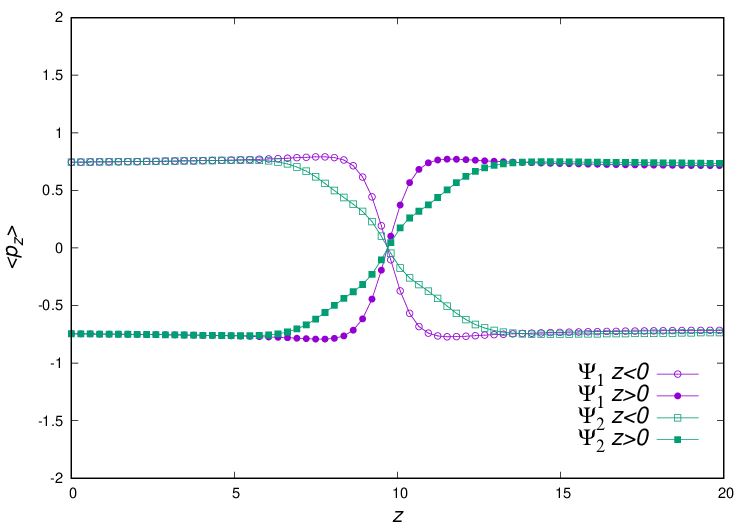}
\caption{Solitonic case of a two states - two states encounter. We show snapshots of the interaction in the high velocity regime of two configurations with two states. In this case the two wave functions show an interference patter. Also shown are the values of $\langle p_z \rangle$ calculated with $\Psi_1$ and $\Psi_2$ that shows the momentum is being transferred from one half domain to the other one. The evolution was calculated on the domain $r\in [0,30],~ z\in[-30,30]$ with resolution $\Delta r= \Delta z = 0.12$ and the configurations were initially located with  $z_0=15$.}
\label{fig:case22solitonic}
\end{figure}

{\it High velocity case.} We again use $p_z=1.5$ and look for a type of solitonic behavior of the system. The results are shown in Fig. \ref{fig:case22solitonic}. At the moment of collision there is an interaction pattern of the two densities $\rho_1$ and $\rho_2$. The solitonic behavior is actually very poor, since the density of the ground state deforms significantly as seen by its amplitude, and the excited state density deforms significantly after the interaction as can be seen in the snapshot at $t=19.6$, where the nodes have been lost. 
%ANA: Los estados permanecen ortogonales durante la evoclucion...

\begin{figure}
\centering
\includegraphics[width= 4.25cm]{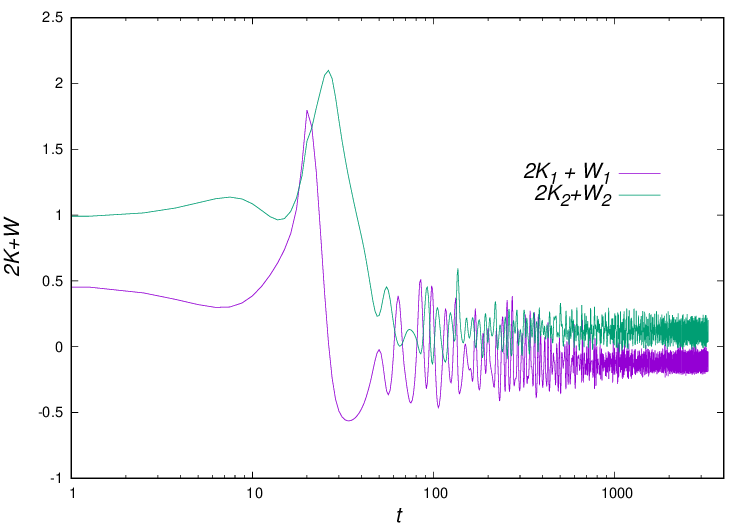}
\includegraphics[width= 4.25cm]{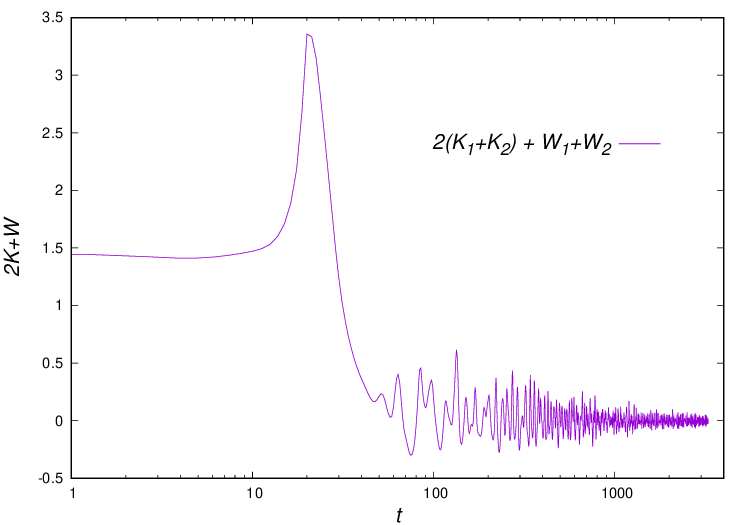}
\includegraphics[width= 4.25cm]{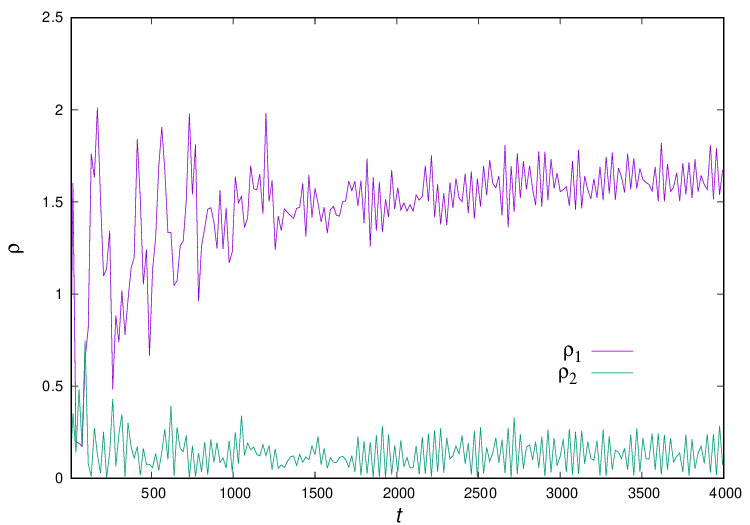}
\includegraphics[width= 4.25cm]{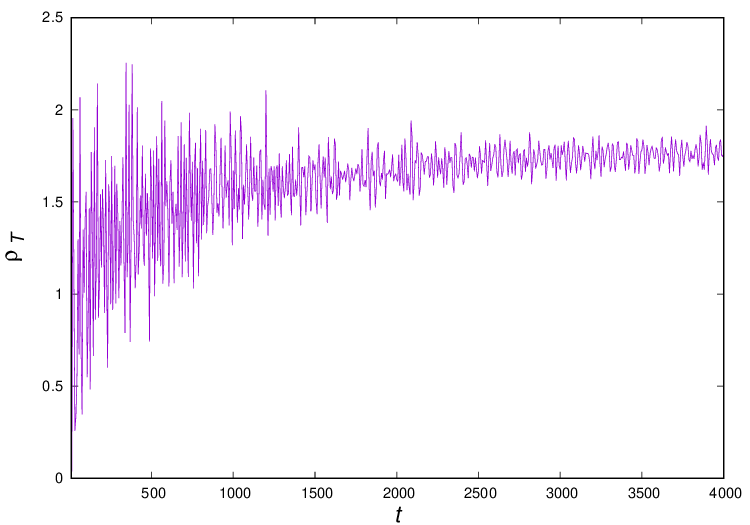}
\caption{Merger case of a two states - two states encounter. In the first panel we show $2K_1+W_1$ and $2K_2+W_2$ as function of time. In the second panel we show the addition of the two, which indicates that the whole system is approaching a virialized state in asymptotic time. We also show the central value of densities $\rho_1$ and $\rho_2$ of the final configuration; the behavior is oscillatory and interestingly does not show the tendency to a constant time independent value., instead it oscillates with various modes. However the central value of $\rho_T=\rho_1+\rho_2$ approaches better to an asymptotic value. The evolution was calculated on the domain $r\in [0,30],~ z\in[-30,30]$ with resolution $\Delta r= \Delta z = 0.12$ and the configurations were initially located with  $z_0=15$.}
\label{fig:case22merger}
\end{figure}

{\it Low velocity case.} For the merger case we use $p_z=0.5$ and show the results in Fig. \ref{fig:case22merger}. We find in the first place that the quantity $2K_i + W_i$ calculated for each state does not approach zero if calculated independently, however the quantity $2(K_1+K_2)+W_1+W_2$ approaches zero, which indicates the system tends toward a virialized state. In the second place we notice that the central density of the final configuration of each of the two states oscillates with no decreasing amplitude. In fact the excited state  sometimes has nodes and some others it does not, a behavior that shows to be periodic. Instead of decreasing, the central value of $\rho_1$ and $\rho_2$ oscillate with various frequency modes and non-decreasing amplitude. This behavior is puzzling because on the one hand the system approaches a virialized state whereas the densities do not seem to approach a stationary regime. Nevertheless, it happens that the addition of the densities of the two states oscillates with a decreasing amplitude.

We choose this particular case to discuss some potential interesting results to the axionic dark matter models. In the context of ultralight axion dark matter, 3D simulations show that the mass function of structures have a profile given by a core with a density profile similar to that of ground state configurations and a halo  surrounding the core \cite{BernalGuzman2006a,BernalGuzman2006b,Schive:2014,Schwabe:2016}.

The idea is to  extract  the density profile of the final configuration in order to estimate potential implications to axion dark matter astrophysics. The final configurations are not stationary, and actually they have never been shown to be either in 2D  \cite{Schwabe:2016} nor in 3D \cite{Schive:2014} analyses. In our case we notice a low frequency mode that modulates the oscillation of the two densities with a period of nearly 500 in code units. What we do is to calculate an average of the total density $\rho_T=\rho_1+\rho_2$ and fit a density profile for such average. For this we averaged snapshots of $\rho_T$ from $t=$3500 to 4000, which is nearly the period that modulates the oscillation of the  densities.

\begin{figure}
\centering
\includegraphics[width= 8cm]{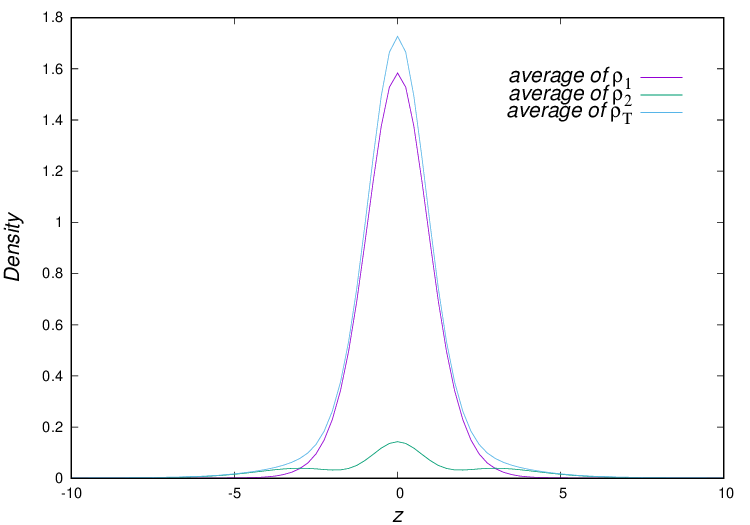}
\caption{Average of $\rho_1$, $\rho_2$ and the total density $\rho_T=\rho_1+\rho_2$ in the time range between 3500 and 4000. This plot shows the contribution of each state to the total density. }
\label{fig:case22merger_avg}
\end{figure}

In Fig. \ref{fig:case22merger_avg} we show the average of the densities of each of the two states and their addition. This average shows the contribution of a core and a halo. This is one of the convenient features of  multi-state configurations \cite{UrenaBernal:2010}. We show here that similar states can be formed through the collision of two configurations with more than one state.

\begin{figure}
\centering
\includegraphics[width= 8cm]{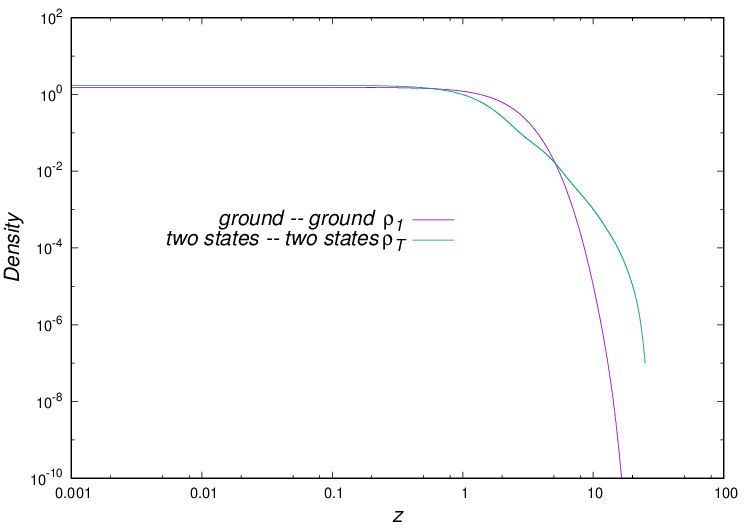}
\caption{Density profile showing a core-like slope and a power law tail. We show three cases, the result of the ground-ground collision case, the result of a ground state with a two state configuration and the result of the merger of two configurations with two states.}
\label{fig:case22merger_slope}
\end{figure}

What we do next is to fit the average profile of $\rho_T$ as done for the ground - ground merger case in section in \ref{subsec:single} and compare. We find that there is a core governed by a Gaussian and a power-law tail given by the profiles%
\begin{eqnarray}\nonumber
\rho_{core}&\sim&e^{-\left(\frac{z}{1.33}\right)^2},\qquad z<z_c\\
\rho_{halo}&\sim& z^{-3.1},\qquad z>z_c
\label{eq:finalProfile2merger}
\end{eqnarray}
\noindent which is illustrated in Fig.  \ref{fig:case22merger_slope}. We define the core-limit $z_c$ as the point where the profile $\rho_T$ has an inflexion. 
This shows the ground state vs ground state head-on mergers gives rise to solitonic solutions resulting profiles, whereas the merger  of two-states vs two-states case results in a core with a tail envelope. 

%

% ----------------------------------------------------------------
% ----->     Final comments     <-----
% ----------------------------------------------------------------
\section{Conclusions and final comments}
\label{sec:final}

We present the head-on collision of multi-state equilibrium configurations ruled by the GPP system. As an attempt to study solitonic behavior of colliding multi-state configurations, we performed simulations with a high head-on momentum, and found that there is no such a strict solitonic behavior as given by definition in terms of the constancy of the density profile. Instead, the coupling between the wave functions of each state through the common gravitational potential, deforms the initial density profiles of each state after the collision in the asymptotic time, in comparison with the case of the collision of two ground state configurations in \cite{Choi2002,BernalGuzman2006b}.

We also simulated the head-on collision with smaller initial momentum in order to produce mergers. In this case some interesting findings turn up. These results are generic taking into account the various cases treated in our analysis. In the first case we analyzed {\it the merger of two ground state configurations} and find that:
1) The resulting configuration relaxes, 2) it tends towards a stationary regime while oscillating,  3) its density profile approaches that of a stationary ground state configuration. This is consistent with the attractor nature of equilibrium configurations. From a naive standpoint, the previous findings might be controversial with the results in \citep{Schwabe:2016}, where the resulting configurations have a core-tail profile. However such controversy would be apparent because in the later case the initial configuration that eventually merge, have non-vanishing orbital momentum and therefore the final configuration has a non-trivial rotation velocity field, whereas our merger involves head-on encounters that lead to a final configuration with non-rotational velocity field.

Another interesting case is {\it the merger of two-state configurations}, for which we find that: 
1) The quantity $2(K_1+K_2)+W_1+W_2$ approaches to zero asymptotically in time, which indicates that the total system tends to a virialized state.
2) A long while after the collision, even though the densities associated with the two states $\rho_1$ and $\rho_2$ oscillate with a modulated amplitude, the total density $\rho_T=\rho_1+\rho_2$ approaches a stationary state. This is an interesting hint concerning to structure formation since, at least in this case when multistates merge, individual configurations do not virialize, however the system does as a whole. 
3) We obtained an averaged profile of $\rho_T$ and found that it has a 
core plus a tail structure that could serve to explain the results in simulations of  structure formation. These results are qualitatively consistent with those of \citep{Schive:2014} and \cite{Schwabe:2016} although having different fits for the core, nonetheless the shape of the tail envelope coincides in either works. This merger case of multi-state configurations is the most illustrative of our analysis, showing the potential of assuming the condensate to be made of multi-states.

So far, we have presented the results of a theoretical study, and have shown how the  densities resulting from the merger of multi-state configurations can be potentially interesting within the frame of the ultralight scalar field dark matter. A direct application of the results shown in this paper within the astrophysical scenario, will have to consist of a massive set of simulations between multi-state configurations, with different linear momentum and relative masses, as an attempt to define a universal density profile resulting from mergers of different initial configurations, assuming the GPP system models the dynamics of a Bose-Einstein condensate of ultralight bosons with masses of the order of $10^{-22}eV$.

%We consider that the multi-state approach is a promising at looking for a universal profile of BEC dark matter structures, together with the possibility of universal profiles of rotating configurations, which would be more consistent with the mergers in 
%\citep{Schwabe:2016} and shown to have interesting profiles and rotation curves \citep{GuzmanRivera2010,GuzmanLora2015}.

% ----->     ACKNOWLEDGMENTS     <-----

\section*{Acknowledgments}
We specially thank Argelia Bernal, for providing the data for the multi-state configurations in \citep{UrenaBernal:2010}. This research is supported by grants CIC-UMSNH-4.9 and CONACyT 258726 (Fondo Sectorial de Investigaci\'on para la Educaci\'on). A.A.L acknowledges financial support from CONACyT posdoctoral fellowship. The simulations were carried out in the computer farm funded by CONACyT 106466 and the Big Mamma cluster at the IFM.

% -------------------------------------------------------
% -----     REFERENCES     ----------
% -------------------------------------------------------

\bibliographystyle{plain}
\bibliography{sample}

\end{document}